\documentclass{desyproc}

\begin{document}
\title{Impact of Axions on the Minimum Mass of Core Collapse Supernova Progenitors}

\author{{\slshape Inma Dom\'\i nguez$^1$, Maurizio Giannotti$^{2}$, Alessandro Mirizzi$^{3}$, Oscar Straniero$^{4}$}\\[1ex]
$^1$ Univ. de Granada, Granada, Spain\\
$^2$ Univ. of Barry, Miami Shores, Florida, USA\\
$^{3}$ Univ. of Bari \& INFN-Sezione di Bari, Italy\\
$^{4}$ INAF-Osservatorio Astronomico d'Abruzzo, Teramo, \& INFN-Laboratori Nazionali del Gran Sasso, Assergi, Italy}


\confID{13889}  
\desyproc{DESY-PROC-2017-XX}
\acronym{Patras 2017} 
\doi  

\maketitle

\begin{abstract}

In this study we include axions in stellar evolution models adopting the current stringest constraints for their coupling to photons 
 and electrons. We obtain that the  minimum stellar mass of Core Collapse Supernova (CCSN) progenitors  is shifted up by nearly 2 M$_\odot$. 
 This result seems to be in tension with the observationaly derived minimum mass of CCSN progenitors. 
\end{abstract}

\section{Introduction}

Stars are known to be good laboratories for particle physics. Axions and Axions Like Particles (ALPs) may 
 be produced in stellar interiors and freely escape, carrying energy out and thus, modifying stellar evolution.  

Axions are weakly interactive particles that were introduced to explain the absence of CP violation in the strong interactions  ~\cite{pe77,we78, wi78}. Later on, these particles and, in general, ALPs  were proposed 
as dark matter candidates ~\cite{si08}. In well motivated axion models, like the DFSZ  ~\cite{di81, zi80}, axions couple to photons and fermions. These couplings are characterized by the corresponding 
  coupling constants, $g_{a\gamma}$ and  $g_{ae}$, being the energy loss rates in these interactions proportional to the square of the coupling constants. In stellar interiors, ALPs that couple to photons are expected to be produced by the Primakoff process and, if they couple to 
 electrons, mainly by the Compton and Bremsstrahlung processes (see ~\cite{ra90}). 

The ALP stellar approach consists of constraining the coupling strengths in order to avoid conflicts with astronomical observations. Moreover, there are also astronomical observations that could be better explained considering an extra-energy sink (see i.e.  ~\cite{gi17}). Among them, the observed decrease of the pulsational period of some white dwarfs ~\cite{is92,co12} and the shape of the observed WD luminosity function  ~\cite{is08,mi14}. In both cases the derived limits are for  $g_{ae}$.

In this work we study the influence of axions on the value of the minimum stellar mass that experiences central carbon burning, Mup (~\cite{be80}) and so, the minimum possible 
 stellar mass that may produce a CCSN. Among CCSNe, type II-P are those expected to come from the evolution of single stars. Their bright supergiant progenitors are identified on the corresponding 
  images taken at the SN locations before the explosion, and the progenitor masses estimated from the luminosities of these supergiants. Masses in the range  7-10 M$_\odot$ have been estimated 
   for the type II-P SNe 2003gd, 2005cs, 2009md, 2006my, 2012A and 2013ej (see  ~\cite{sm15}), while theoretical predictions for the minimum masses of CCSNe are above  
   9-10 M$_\odot$. As we will show, the presence of axions coupled to photons and electrons would increase the progenitor mass expected by these current prediction, possibly beyond observational constraints. 

\section{Numerical simulations \& Results} 

We consider DFSZ axions produced through the mentioned processes (Primakoff, Compton and Bremsstrahlung), adopting for the coupling constants the most updated   
  limits that are also provided with corresponding uncertainties: 

\begin{itemize}

\item $g_{ae} \le 4.3\times 10^{-13}$  at 95$\%$ CL ~\cite{vi13}, based on the luminosity of the RGB tip derived from globular clusters (GCs). 

\item $g_{a\gamma} \le 0.66\times 10^{-10}$ GeV$^{-1}$ at 95$\%$ CL, based on the 
 R parameter derived for GCs ~\cite{ay14,st15}, and being also the the experimental upper limit obtained by 
 the CAST collaboration ~\cite{ca17}.  

\end{itemize}

\begin{figure}[hb]
\centerline{\includegraphics[width=0.45\textwidth]{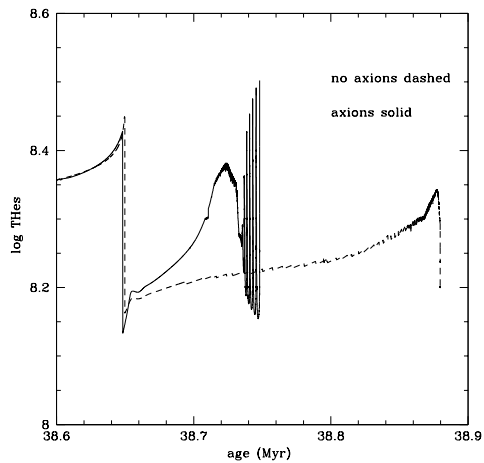}}
\caption{Evolution in time 
 of the maximum temperature within the He-shell for the reference model of 7.5 M$_\odot$, without axions (dotted line), and for a model with the same mass, including axions (solid line).}\label{Fig:THes}
\label{sec:figures}
\end{figure}

The FUNS code ~\cite{st06,cr09,cr11}, modified to include the corresponding axion processes, is 
  used for all the numerical simulations. The axion energy loss rate for the Primakoff is taken from ~\cite{ra90}, for the Compton from 
 ~\cite{ra95}, for the non-degenerate Bremstrahlung from ~\cite{ra95} and for the 
  degenerate Bremsstrahlung from ~\cite{na87,na88}. Rates and interpolations have been revised by the authors.  

We compute models in the mass range 7 to 11 M$_\odot$, with a resolution of 0.2 M$_\odot$, and assuming 
  solar initial chemical composition, Z= 0.013 \& Y= 0.26.  All models are 
 computed from the pre-MS to central C-burning, or alternatively, to CO core cooling (along the AGB phase). 

We have first identified, as a reference, the minimum initial stellar mass needed for central C-burning, Mup, without axions, 
  and it turns out 
 to be 7.5 M$_\odot$. Then, we compute models  including the previous axion energy 
  loss rates and we find that models in the mass range 7.5 to 9.1 M$_\odot$ do not ignite carbon, being Mup 9.2 M$_\odot$. 

The main reason of this shift of Mup by  1.7 M$_\odot$ is the smaller CO core developed, for the same initial stellar mass, by models 
 that include axions. The growth of the CO core, during the so called early-AGB phase, is halted earlier because the evolution is faster: nuclear energy has to compensate 
 the axion energy losses within the He-shell. This effect was discussed by some of us, in another context, some years ago (~\cite{do99}). The different time-scales are clearly seen in Fig. \ref{Fig:THes}. The early-AGB time is reduced  by the axion effect, from 2.223$\times 10^{5}$ yrs
 to 7.76$\times 10^{4}$ yrs  wich results in final CO core masses of  1.066 M$_\odot$ for the reference model, and 0.937 M$_\odot$ for the model with axions.  

We identify both, Compton and Primakoff, as being the most important processes and equaly relevant. For example, when only Primakoff is included Mup is 8.4 M$_\odot$, so Mup is shifted up, with respect to models 
  without axions by 0.9 M$_\odot$ instead of by 1.7 M$_\odot$ when Compton is also included.   

However, the change in the CO core mass for a given total stellar mass is not the only factor that influences Mup. We find that an increase of the final CO core mass by 0.113 M$_\odot$, with respect to that of the reference model, is needed for C-burning.  Axions cool the degenerate inner part of the CO core and this is due the Bremmsstrahlung process. In fact, when Compton and Primakoff are included but 
 Bremmsstrahlung is not included, Mup is 9.0 M$_\odot$ instead of 9.2 M$_\odot$.

\section{Conclusions} 

The production of DFSZ axions in stellar interior, assuming the current stringest upper limits for the coupling constant values, represents an important energy sink. 
 In this work we have focused on the evolution of stellar masses that are close to the observed minimum progenitor mass for CCSNe. Assumed values for  
 $g_{ae}$ and $g_{a\gamma}$ are    
  4.3$\times 10^{-13}$ 
   and 0.66$\times 10^{-10}$ GeV$^{-1}$, respectively. Our main result is that the minimum stellar mass that experiences central carbon burning, Mup, is shifted up 
  by 1.7 M$_\odot$, from 7.5 M$_\odot$ for the reference model to 9.2 M$_\odot$ when axions are included, implying that: 

\begin{itemize}
\item Stars with masses smaller than 9.2 M$_\odot$,  would produce CO WDs, potential progenitors of Type Ia Supernovae (SNIa). 
 This may help to solve the problem related with the predicted low theoretical SNIa rates as compare with the rates derived from observations (see ~\cite{ma14} for a review). 
\item CCSNe would only be produced by stars with masses greater than 9.2 M$_\odot$. In fact, considering that a mass interval over Mup is expected to 
  produce ONe WDs and electron capture SNe, the minimum mass for CCSNe progenitor would be even greater. This would be  in conflict with the observationaly derived minimum  
  progenitor masses of CCSNe,  7-10 M$_\odot$  ~\cite{sm15}, . 
\end{itemize}

It is remarkable that the next generation of dedicated axion experiments, like ALPSII ~\cite{ba13} and IAXO ~\cite{gi16}, will focus on the range of parameters that is being tested by stellar models, 
 like those considered in this work.

\section{Acknowledgments}

I.D ackknowledges founding from the MINECO-FEDER grant
AYA2015-63588; A.M. is supported by the MIUR and INFN through the Theoretical Astroparticle Physics project; 
O.S. acknowledges founding from the PRIN-MIUR grant 20128PCN59.

\section{Bibliography}


\begin{footnotesize}

\end{footnotesize}


\end{document}